\documentclass[10pt,a4paper,twocolumn,english,pra,aps,showpacs,floatfix,groupedaddress,superscriptaddress]{revtex4}

\usepackage{graphicx}
\usepackage{epsfig}
\usepackage[english]{babel}
\usepackage{amsmath}
\usepackage{amssymb}
\usepackage{amsfonts}
\usepackage{longtable}
\usepackage{dcolumn}
\usepackage{bm}

\begin{document}
\title{Optimization of Short Coherent Control Pulses}

\author{S. Pasini}
\affiliation{Lehrstuhl f\"{u}r Theoretische Physik I,  Technische Universit\"{a}t Dortmund,
 Otto-Hahn Stra\ss{}e 4, 44221 Dortmund, Germany}

\author{T. Fischer}
\affiliation{Lehrstuhl f\"{u}r Theoretische Physik I,  Technische Universit\"{a}t Dortmund,
 Otto-Hahn Stra\ss{}e 4, 44221 Dortmund, Germany}
\author{P. Karbach}
\affiliation{Lehrstuhl f\"{u}r Theoretische Physik I,  Technische Universit\"{a}t Dortmund,
 Otto-Hahn Stra\ss{}e 4, 44221 Dortmund, Germany}

\author{G. S. Uhrig}
\affiliation{Lehrstuhl f\"{u}r Theoretische Physik I,
 Technische Universit\"{a}t Dortmund,
 Otto-Hahn Stra\ss{}e 4, 44221 Dortmund, Germany}

\date{\rm\today}

\begin{abstract}
The coherent control of small quantum system is considered. 
For a two-level system coupled to an arbitrary bath we consider
a pulse of finite duration.  We derive the leading and the next-leading order 
corrections to the evolution operator due to the non-commutation of the pulse 
and the bath Hamiltonian. The conditions are computed that make the leading 
corrections  vanish. The  pulse shapes optimized in this way are given for
$\pi$ and $\frac{\pi}{2}$ pulses. 
\end{abstract}


\maketitle

\section{Introduction}
For a long time, the coherent control of quantum systems
has been an important issue. For instance, nuclear magnetic resonance (NMR)
is a widely used technique for the analysis of chemical compounds and
for imaging techniques. Another broad field of application is the coherent
control of the state of quantum dots, both in the charge and in the spin 
degrees of freedom. In recent years, however, coherent 
control has gained an even more vivid interest because it is the indispensable
prerequisite in quantum information processing. 

In its most basic form, the quantum state of a single quantum bit (qubit)
is to be changed in a given way. This is called a single qubit gate.
A qubit is a two-level system which can be conveniently seen as $S=1/2$ system.
So we use this spin language to describe the qubit gates. The two most
common gates are the $\pi$ pulse, which flips the spin from up to down and
vice-versa, and the $\pi/2$ pulse or Hadamard gate, which rotates the spin
by $90^\circ$ away from the $S_z$ direction.

The $\pi$ pulse is particularly interesting for dynamic decoupling
\cite{viola98,ban98,facch05,khodj05,cappe06,witze07a,yao07,uhrig07}. 
In a nutshell,
this technique aims at decoupling the single qubit from its environment
as well as possible by switching the qubit state by single qubit
gates. Mostly, the $\pi$ pulse is considered. 
The idea comes from the spin-echo technique 
in NMR where a $\pi$ pulse is used to refocus
the precessing magnetization. In NMR, the use of sophisticated protocols
of $\pi$ and $\pi/2$ pulses is common to suppress unwanted couplings
between nuclear spins \cite{haebe76,vande04}.

Experiments using $\pi$ pulses have obtained many encouraging results in 
prolonging the  coherence time of a qubit 
\cite{morto06,petta05,greil06a,greil06b}. More complex
pulse sequences based on $\pi/2$ pulses have also proved useful by deminishing
the rate of decoherence in quantum registers considerably \cite{kroja06}.

In the theoretical treatments the generic ingredient to dynamic
decoupling is an idealized pulse, mostly a $\pi$ pulse, which is 
considered to be instantaneous. This means that its amplitude is
infinite in the sense of a $\delta$-function. This assumption
is convenient because the Hamiltonian $H_0$ of the pulse and the Hamiltonian
 $H$ of the system at rest do not commute in general. The Hamiltonian
$H$ comprises the Hamiltonian of the qubit itself, the Hamiltonian
of the environment (henceforth called ``bath'') and the coupling between
the qubit and the bath which is responsible for its decoherence.
But during the pulse $H_0$ dominates and $H$ is negligible and for the rest of
the time only $H$ is present. Hence it is relatively straightforward to deal 
with the time evolution. 

Experimentally, however, this idealized situation 
is not realistic. The real pulses are always of finite amplitude and
of finite length. So the question has to be addressed which effect a
finite pulse duration $\tau_p$ has. We want to elucidate
this issue in the present work theoretically in a fully quantum 
mechanical framework.

The issue of pulses of finite duration and finite amplitude
has been considered in other papers. Viola and Knill generalized
their previous approach of averaging over a symmetry group \cite{viola99a}
by instantaneous changes of the effective Hamiltonian to continuous
changes requiring only  finite, bounded control amplitudes \cite{viola03}.
But the presence of the coupling to the bath \emph{during} the pulses 
has not been considered.

Khodjasteh and Lidar considered the effect of finite pulse duration
in a comprehensive comparison of various schemes for $\pi$ pulse sequences
\cite{khodj07}. They did not consider to make the detrimental effects
of the finite pulse duration vanish on the level of the individual pulse,
but they discussed possible cancellations on the higher hierarchical level of 
the pulse sequence. In the present work, we will choose another route
and aim at reducing the decoherence due to finite pulse duration already
on the level of the single pulse.
As a side remark, we note that the statement in Ref.\ \onlinecite{khodj07} 
on  optimized pulse sequences \cite{uhrig07}
that they hold only for bosonic bath operators is very likely 
not to be correct. It has been conjectured very recently by Lee et al.\ 
\cite{lee07} that the optimized sequences apply to the most general phase 
decoherence model. 

Very recently, the effect of \emph{classical} random telegraph noise 
during pulses of finite length has been discussed numerically \cite{motto06}.
It was shown that shaping the pulse amplitude in a particular way
can be used to improve its performance. This means that one can
optimize the pulse such that its effect is closer to the desired one,
e.g.\ a $\pi$ pulse.

Tuning pulses by shaping them is a possibility of optimization 
which has been discussed intensively in the vast literature on
NMR. But the objectives were mostly different from the 
consideration of the effects of a bath. For instance, Geen and Freeman
aimed at frequency selectivity \cite{geen91}, i.e.\ the pulse should act only 
on spins on or close to resonance, but leave others unchanged.
Closest to our scope are the investigations by Tycko \cite{tycko83},
Cummins et al.\ \cite{cummi00,cummi03}, and Brown et al.\ \cite{brown04}
where composite pulses are optimized to be robust against
off-resonance effects and pulse length inaccuracies. Again, no dynamic but
only static effects are considered. Interestingly, we will see that
some of the pulse shapes fulfilling the conditions in first order derived here
for fully dynamical baths coincide with pulse shapes previously
derived to compensate static effects.

In this work, we present a systematic expansion in the pulse duration
$\tau_p$ considering a qubit coupled to a fully quantum mechanical and
dynamical  bath. 
This means that the related energy scale $\omega_p:=1/\tau_p$ 
($\hbar$ is set to unity for simplicity) is considered to be very
large so that all other energy scales, namely the coupling $\lambda$
between qubit and bath and the internal energy scale of the bath
$\omega_b$, are small relative to $\omega_p$. We expand in 
$\lambda/\omega_p =\tau_p\lambda$ and  $\omega_b/\omega_p =\tau_p\omega_b$.
The zeroth order is the instantaneous pulse. 
We compute the first and the second order and discuss the conditions
for which they vanish. These conditions allow us to
predict pulse shapes which approximate instantaneous pulses in spite
of their finite pulse duration.

We emphasize that it is not our primary aim to eliminate the coupling
between qubit and bath. On the contrary, it turns out
that the expansion around the instantaneous pulse keeps this coupling.
But the coupling to the bath  is disentangled from the actual pulse.
They are expressed by separate, subsequent time evolution operators.
Given the thus optimized pulse shapes,
the coupling to the bath can be compensated on the higher hierarchical
level of an appropriate pulse \emph{sequence}, i.e., by dynamic decoupling
\cite{viola98,ban98,facch05,khodj05,cappe06,witze07a,yao07,uhrig07}.

The paper is organized as follows. After introducing the model (Sect.\ II), 
we start describing the method we used to derive our general equations for the 
first and second order  (Sect.\ III). Then we discuss the results and
provide  various examples for pulses that satisfy our conditions (Sect.\ IV). 
At last we  discuss our results and compare them with other proposals for 
shaped pulses  in the literature (Sect.\ V).

\section{Model} 

\begin{figure}
\begin{center}
     \includegraphics[width=\columnwidth]{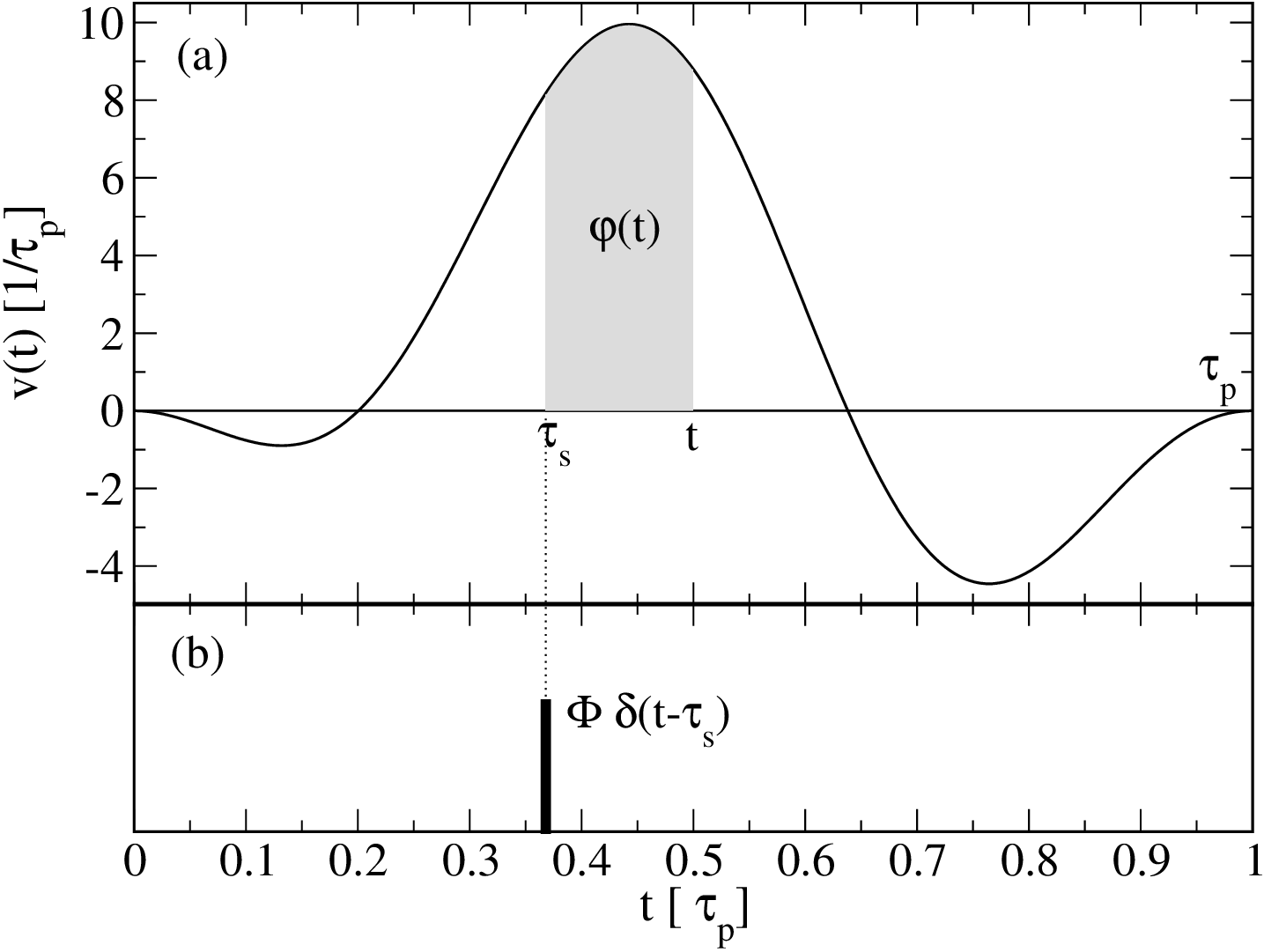}
\end{center}
\caption{(a) Plot of a generic pulse $v(t)$. 
The x- and y-axis values are expressed in units of
$\tau_p$ and $1/\tau_p$, respectively. The shaded region represents an 
example for $\varphi(t)$, that is the integrated pulse amplitude
between the instant $\tau_s$ and a generic instant $t$ (see text).\\
(b) Sketch of the approximately equivalent idealized instantaneous
pulse.
} 
\label{fig1}
\end{figure}

We consider the time evolution operator from $t=0$ to $t=\tau_p$
\begin{equation}
\label{ev_op} 
U_p(\tau_p ,0)=\text{T}\left\{\exp\left(-i \int_{0}^{\tau_p} 
H_0(t) dt -i H \tau_p \right)\right\}\ .
\end{equation} 
For simplicity the pulse is chosen to start at $t=0$;  T stands for the 
usual time-ordering. The control Hamiltonian of the pulse shall be given by
\begin{equation}
\label{hamilt_pulse} 
H_0=  v(t) \sigma_y,
\end{equation} 
where $v(t)$ stands for the pulse shape as function of time, see Fig.\ 
\ref{fig1}. The time-independent Hamiltonian of the bath and the coupling
to the qubit reads
\begin{equation}
\label{hamilt_gen_bath} 
H=H_b + \lambda A \sigma_z,
\end{equation} 
with $H_b$ representing a generic bath and $A$ its coupling operator 
to the qubit. The internal energy scale of $H_b$ shall be denoted
by $\omega_b$. Note that $H_0$ and $H$ do not commute so that
the evolution operator (\ref{ev_op}) is a non-trivial expression.

The  model \eqref{hamilt_gen_bath} 
does not include spin flips; hence it implies an 
infinite life time $T_1$. But the decoherence of a precessing
spin in the $xy$ plane is described in full generality because
we do not specify for which operator $A$ stands and the bath
dynamics described by $H_b$. Such a model is experimentally 
very well justified as the effective model in the limit of 
a large applied magnetic field which implies that  other
couplings between the quantum bit spin and the bath are
averaged out, see for instance Ref.\ \onlinecite{li07,li08a}.

As an example only one may  consider the spin-boson model
\begin{equation}
\label{sBM}
H=\sum_i \omega_i b^\dagger_i b_i+\sigma_z
\sum_i \lambda_i (b^\dagger_i + b_i)
\end{equation} 
with $\lambda:= \max_i \lambda_i$, 
$A=\sum_i (\lambda_i/\lambda) (b^\dagger_i + b_i)$
and $\omega_b:= \max_i \omega_i$.

The primitive of the pulse amplitude is
\begin{equation}
\label{phi-def}
\varphi(t):=\int_{\tau_s}^t v(t')dt',
\end{equation}
where the integration starts from an instant $\tau_s \in [0,\tau_p]$.
The intended angle of rotation is given by twice the total area
under the curve $\theta:=2\Phi=2(\varphi(\tau_p)-\varphi(0))$.
This area has to be kept constant while investigating the limit
$\tau_p\to 0$ so that the amplitude $v(t)$ has to tend to infinity
as $v(t)\propto 1/\tau_p$. To simplify our notation later, we define
also $\psi(t):=2\varphi(t)$.

The aim is to replace the real pulse given by $v(t)$ by an approximately
equivalent instantaneous pulse acting at the instant $\tau_s$, 
see Fig.\ \ref{fig1}. The equivalence shall hold up to linear order
or quadratic order in $\tau_p$. We highlight that the
time evolution before and after the equivalent instantaneous pulse
shall be governed by $H$, i.e., including the decohering interaction between
qubit and bath. This shall be accounted for on the higher hierarchical
level of appropriate pulse sequences.

\section{Method and General Equations} 

\subsection{General Equation}

Let $\tau_s$ be an arbitrary instant inside the interval
$[0,\tau_p]$ at which we want to split the time evolution
\begin{equation} 
\label{unitary-split}
U_p(\tau_p ,0)=U_p\left(\tau_p ,\tau_s\right)U_p\left(\tau_s ,0\right).
\end{equation}
The guiding idea is shown in Fig.\ \ref{fig1}b. We want to replace the
full evolution by a first part during $[0,\tau_s]$ 
where only the Hamiltonian $H$ is active. Then the instantaneous
pulse shall follow. Finally, another intervall $[\tau_s,\tau_p]$ of
the dynamics governed by $H$ 
terminates the pulse. This motivates the two following
ans\"atze for the two unitary operators in Eq.\ \ref{unitary-split}
\begin{subequations}
\label{ansatz}
\begin{eqnarray}
\label{ansatz1}
U_p\left(\tau_p ,\tau_s\right) &=& e^{-i (\tau_p-\tau_s)H } \
e^{-i\sigma_y \int_{\tau_s}^{\tau_p}
 v(t) dt} \ U_2\left(\tau_p ,\tau_s\right)\qquad
\\
\label{ansatz2}
 U_p\left(\tau_s ,0\right) &=& U_1\left(\tau_s ,0\right) \ e^{-i\sigma_y
\int_{0}^{\tau_s}v(t)dt} \ e^{-i \tau_s H},
\end{eqnarray} 
\end{subequations}
which \emph{define} the operators $U_1\left(\tau_s ,0\right)$ and 
$U_2\left(\tau_p,\tau_s\right)$. Both, $U_1$ and $U_2$ must be seen
as the corrections which are necessary because the left hand sides
in Eqs.\ \ref{ansatz} do not equal the right hand sides without them
since the pulse and the bath Hamiltonians do no commute.
Note that no time-ordering is necessary for the exponential of
$H$ because it is constant. For the exponential of $H_0$ no
time-ordering is necessary despite the time dependence because
the commutator $[H_0(t),H_0(t')]=0$ for any two times $t$ and $t'$.

If the corrections are small, e.g.\ $U_2(\tau_p,\tau_s)U_1(\tau_s,0)=1+
{\cal O}(\tau_p^2)$ up to linear order in $\tau_p$, this implies
\begin{equation}
U_p(\tau_p ,0)= e^{-i (\tau_p-\tau_s)H } \
e^{-i\sigma_y \int_{0}^{\tau_p}
 v(t) dt} e^{-i \tau_s H } 
\end{equation}
which corresponds to the ideal, instantaneous pulse sketched in Fig.\
\ref{fig1}b 
at $\tau_s$ while the dynamics before and after $\tau_s$
includes both the bath dynamics and the coupling between qubit
and bath.

The unitary corrections $U_1$ and $U_2$ are determined from the 
Schr\"{o}dinger equation.
We start analyzing $U_2$ (\ref{ansatz1}).  Let $\tau$
be a time instant in the interval $\left]\tau_s,\tau_p\right]$ such
that
\begin{equation}
\label{opev_2_tau} 
U_p\left(\tau ,\tau_s\right)=e^{-i \Delta\tau H}
\ e^{-i \int_{\tau_s}^{\tau}
 H_0(t) dt} \ U_2\left(\tau ,\tau_s\right),
\end{equation}
where we use the difference
\begin{equation}
\label{eq:delta}
\Delta x:=x-\tau_s
\end{equation}
 generally for any variable $x$; here in particular for
$x=\tau$.
The Schr\"{o}dinger equation  of $U_p\left(\tau ,\tau_s\right)$ 
is the usual one
\begin{equation}
\label{schr_eq_2} 
i\partial_\tau U_p\left(\tau
,\tau_s\right)=(H_0(\tau) + H )\ U_p\left(\tau ,\tau_s\right). 
\end{equation} 
Inserting Eq. (\ref{opev_2_tau}) into (\ref{schr_eq_2}) one obtains
\begin{eqnarray}
\label{deriveU2}
\nonumber 
&& \hspace*{-4mm}
\left\{H e^{-i \Delta\tau H }  + e^{-i \Delta\tau H }H_0(\tau)
\right\} 
e^{-i\int_{\tau_s}^{\tau} H_0(t) dt}  \ U_2\left(\tau ,\tau_s\right)
\\ \nonumber
&& \qquad\qquad +i\ e^{-i \Delta\tau H } \ e^{-i \int_{\tau_s}^{\tau}
 H_0(t) dt} \ \partial_\tau U_2\left(\tau ,\tau_s\right) 
\\ 
&&  = (H_0(\tau) + H )e^{-i \Delta\tau H} \ e^{-i \int_{\tau_s}^{\tau}
 H_0(t) dt} \ U_2\left(\tau ,\tau_s\right).
\end{eqnarray}
It is easy to see that the terms with $H$ cancel so that
the differential equation for $U_2$ becomes
\begin{subequations}
\begin{equation}
\label{diffeq_U2} 
i \partial_\tau U_2\left(\tau
 ,\tau_s\right)=\left( \widetilde {H}_0(\tau)-H_0(\tau)\right)U_2\left(\tau
 ,\tau_s\right),
\end{equation}
where
\begin{equation}
\label{Htilde}
\widetilde{H}_0(\tau):=e^{i \int_{\tau_s}^{\tau} H_0(t) dt}e^{i \Delta\tau H}
H_0(\tau)e^{-i \Delta\tau H } \ e^{-i \int_{\tau_s}^{\tau} H_0(t) dt}.
\end{equation}
\end{subequations}
Formal integration leads to
\begin{subequations}
\begin{equation}
\label{inteq_U2} 
U_2\left(\tau
 ,\tau_s\right) = \text{T}
\left[\exp\left(-i\int_{\tau_s}^\tau F(t)dt\right)
\right], 
\end{equation} 
where
\begin{equation}
\label{F2}
F(t):= \widetilde{H}_0(t)-H_0(t).
\end{equation}
\end{subequations}
From its defintion it is obvious that $F(t)$ vanishes identically if 
$H_0$ and $H$ commute. This is the case if there is no coupling between qubit 
and bath. Closer inspection of Eqs.\ (\ref{Htilde},\ref{F2}) shows that
$F(t) = {\cal O}(t\lambda v(t))$. More generally, $F(t)$ can be expanded in
a series of $H$, i.e.\ in the parameters $t \lambda$ and $t\omega$ which
amounts up to an expansion in $\tau_p$. Thus our approach provides the intended
expansion in the shortness of the pulse.

With the definitions
\begin{equation}
\label{sigma_of_t} 
\sigma_y (\Delta t) := e^{i \Delta t H}\sigma_y e^{-i \Delta t H }
\end{equation}  
and (\ref{phi-def}) the time-dependent operator $F(t)$ can be written
in the compact form 
\begin{equation}
\label{F2_rewritten}
F(t)= v(t)\left[ e^{i\sigma_y \varphi(t)}\
\sigma_y(\Delta t)\ e^{-i\sigma_y \varphi(t)}-\sigma_y \right].
\end{equation} 

Next, we treat $U_1$ by the analogous procedure solving
the Schr\"{o}dinger equation for (\ref{ansatz2}). Let $\tau$ be a
generic instant in the interval $\left[0,\tau_s\right[$, the ansatz
for this interval is
\begin{equation}
\label{opev_1_tau}
U_p\left(\tau_s ,\tau \right)=U_1\left(\tau_s,\tau\right) 
\ e^{i \varphi(\tau)\sigma_y} \ e^{i \Delta\tau H}. 
\end{equation} 
To obtain the Schr\"{o}dinger equation in the second time argument
of $U_1\left(\tau_s,\tau\right)$ it is convenient to consider
the hermitian conjugate of (\ref{schr_eq_2}) yielding
\begin{equation}
\label{schr_eq_1} 
-i\partial_\tau U_p\left(\tau_s ,\tau \right)=
U_p\left(\tau_s ,\tau \right)[H_0(\tau) + H ] 
\end{equation} 
where the property $U_p^\dagger\left(t_1 ,t_2 \right)=U_p\left(t_2 ,t_1
\right)$ has been exploited. The procedure is the same as for $U_2$
with the only difference that now the exponentials
occur on the right side of $U_1$
\begin{equation}
-i\partial_\tau U_1\left(\tau_s ,\tau \right)=
U_1\left(\tau_s ,\tau \right) F(\tau)
\end{equation} 
with the same $F(t)$ as defined in (\ref{F2}).
Formal integration yields
\begin{equation}
\label{inteq_U1} 
U_1\left(\tau_s
 ,\tau\right) = \text{T}
\left[\exp\left(-i\int_{\tau}^{\tau_s} F(t)dt\right)
\right]. 
\end{equation} 
Finally, we combine both corrections $U_1$ and $U_2$ 
\begin{eqnarray}
\nonumber
&& U_p(\tau_p,0) = U_p\left(\tau_p,\tau_s\right)U_p\left(\tau_s,0\right)\\
&&= e^{-i(\tau_p-\tau_s) H} e^{-i\sigma_y \varphi(\tau_p)} U_F(\tau_p,0)
e^{i\sigma_y \varphi(0)} e^{-i \tau_s H},\qquad
\label{tot_ev_op} 
\end{eqnarray}
where
\begin{eqnarray}
\nonumber
U_F(\tau_p,0)&:=&U_2\left(\tau_p,\tau_s\right)U_1\left(\tau_s,0\right)
\\ \nonumber
&=& \nonumber
\mathrm{T} \left\{ e^{-i\int_{\tau_s}^{\tau_p} F(t) dt}\right\}
\mathrm{T} \left\{ e^{-i\int_{0}^{\tau_s} F(t) dt}\right\}
\\ 
&=&
\mathrm{T} \left\{ e^{-i\int_{0}^{\tau_p}
 F(t) dt}\right\}.
\label{U_F}
\end{eqnarray}
Thus the total correction $U_F$ is given by the time-ordered product
of $F(t)$. Thereby, we have derived the general expression for the
difference between the unitary action of the real pulse as sketched
in Fig.\ \ref{fig1}a and the idealized, instantaneous pulse in 
Fig.\ \ref{fig1}b.

\subsection{Expansion to Second Order in $\tau_p H$}

We want to find the conditions under which $U_F(\tau_p,0)$  can be
approximated by the identity operator. Because $F(t)$ is 
${\cal O}(\lambda)$ for $\lambda\to 0$
 the expansion in $F$ is
an appropriate first step. 
A convenient way to obtain the $n$-th order of an expansion in $\lambda$
and $\omega_b$ is to compute up to the $n$-th order of the
Magnus expansion from average Hamiltonian theory \cite{magnu54,haebe76}.
Then the resulting expressions are expanded to $n$-th order
in $\lambda$ and $\omega_b$. The advantage of this approach 
over a direct expansion is that the expansion is done in the
argument of the exponential.

The Magnus expansion reads
\begin{eqnarray}
\nonumber
U_F(\tau_p,0) &=& 
\text{T}\left[\exp\left(-i\int_0^{\tau_p} F(t)dt\right)
\right]
 \\
&=&
\exp\left([-i\tau_p({F}^{(1)}+{F}^{(2)}+{F}^{(3)}+\ldots)\right],\qquad
\label{magnus_formula}
\end{eqnarray} 
where ${F}^{(1)}=\frac{1}{\tau_p}\int_0^{\tau_p}F(t)dt$
is the average value of $F$. The next-leading term ${F}^{(2)}$ comprises the
commutators of $F$ with itself at different times
\begin{equation}
\label{magnus_formula_2}
{F}^{(2)}=\frac{-i}{2\tau_p}\int_0^{\tau_p}dt_1\int_0^{t_1}dt_2[
F(t_1), F(t_2)]. 
\end{equation}

Thus from the Magnus expansion
we know at least the two leading orders in powers of $F(t)$. 
Next, we
expand $F(t)$ itself which is still a complicated quantity, see
Eq.\ (\ref{F2_rewritten}). To obtain the two leading orders in $H$
we expand $F(t)$ in powers of $H$ corresponding to an expansion in
$\tau_p$. We use the identity
\footnote{This identity can be derived by simply expanding the exponential and 
then rearranging the  terms with the same order in $\Delta t$ as a sum of 
nested commutators.}
\begin{equation}
\label{sigma_delta_t}
\sigma_y (\Delta t)=\sigma_y
+\sum_{n=1}^\infty \frac{i^n}{n!}(\Delta t)^n
[[H,\sigma_y]]_n 
\end{equation} 
where $\Delta t$ is used as defined in \eqref{eq:delta}. The
 notation stands for
\begin{equation}
\label{commutators_1_2ord}
\begin{split} 
[[H,\sigma_y]]_1 &=[H,\sigma_y]=-2i \sigma_x \lambda A  \\
[[H,\sigma_y]]_2 &=[H,[H,\sigma_y]]=-2i \sigma_x \lambda [H_b,A]+4\sigma_y 
\lambda^2 A^2\qquad\\
[[H,\sigma_y]]_3 &=[H,[H,[H,\sigma_y]]]=\ldots ,
\end{split}
\end{equation}
to obtain
\begin{eqnarray}
\nonumber
&&\sigma_y ( \Delta t)-\sigma_y =  2 \sigma_x \Delta t \lambda A+\\
&&\qquad\left(i\sigma_x\lambda[H_b,A]
-2\sigma_y\lambda^2 A^2\right)(\Delta t)^2+{\cal O}(\Delta t^3).
\qquad
\label{sigma_minus_sigma_1}
\end{eqnarray} 
Inserting the above expansion in the equation (\ref{F2_rewritten}) of $F(t)$
and using the elementary relation
\begin{equation}
e^{i\sigma_y \varphi(t)}\ \sigma_x\ e^{-i\sigma_y \varphi(t)}
=
\cos(\psi) \sigma_x+\sin(\psi) \sigma_z
\end{equation}
where 
$\psi(t)=2\varphi(t)$ as well as the Magnus expansion (\ref{magnus_formula})
allows us to compute the leading orders of $U_F$ in exponential 
representation
\begin{equation}
U_F(\tau_p,0) = \exp[-i(\eta^{(1)}+\eta^{(2)}+\ldots)]
\end{equation}
with
\begin{subequations}
\begin{eqnarray}
\eta^{(1)} &=&  (\eta_{11}\sigma_x+\eta_{12}\sigma_z)\lambda A\\ 
\eta^{(2)} &=&
i\left(\eta_{21}\sigma_x + \eta_{22}\sigma_z\right)\lambda [H_b,A]
+ \eta_{23} \sigma_y \lambda^2 A^2.\qquad
\end{eqnarray} 
\end{subequations}
Note that $[H_b,A]={\cal O}(\omega_b)$ so that $\eta^{(2)}$ 
truly represents the desired second order in the shortness of the pulse.
The coefficients $\eta_{ij}$ are given by the integrals
\begin{subequations}
\begin{eqnarray}
\label{eta11}
&&\eta_{11} = 2\int^{\tau_p}_{0} \Delta t v(t) \cos(\psi(t)) dt\\
\label{eta12}
&&\eta_{12} = 2\int^{\tau_p}_{0} \Delta t v(t) \sin(\psi(t)) dt\\
\label{eta21}
&&\eta_{21} = \int^{\tau_p}_{0} \Delta t^2 v(t) \cos(\psi(t)) dt\\
\label{eta22}
&&\eta_{22} = \int^{\tau_p}_{0} \Delta t^2 v(t) \sin(\psi(t)) dt\\
\label{eta23}
&&\eta_{23} = -2\int_0^{\tau_p} (\Delta t)^2 v(t) dt\\
&&\quad +4\int_0^{\tau_p}dt_1\int_0^{t_1}dt_2
\Delta t_1 \Delta t_2 v_1 v_2 \sin(\psi_1 -\psi_2),
\nonumber
\end{eqnarray}
\end{subequations}
where we used $\psi_i$ for $\psi(t_i)$ and $v_i$ for $v(t_i)$ 
for $i\in\{1,2\}$ in the last line.
The differences are used as defined in \eqref{eq:delta}. The last 
line of \eqref{eta23} results from the commutator 
(\ref{magnus_formula_2}) yielding
\begin{eqnarray}
\nonumber
\left[F(t_1),F(t_2)\right] &\!\! = \!\! &
\Delta t_1\Delta t_2 v_1 v_2 \left([\sigma_z,\sigma_x]\sin\psi_1\cos\psi_2
\right.\\ \nonumber
&&   \left.+[\sigma_x,\sigma_z]\cos\psi_1\sin\psi_2\right)\lambda^2 A^2\\ \nonumber
&\!\! = \!\!& 2i \Delta t_1\Delta t_2 v_1v_2\sigma_y\times\\
\nonumber
&&  \left( \sin\psi_1\cos\psi_2-\cos\psi_1\sin\psi_2\right)
\lambda^2A^2\\ 
& \!\!= \!\! & 2i \Delta t_1\Delta t_2 v_1v_2\sigma_y
\sin\left(\psi_1-\psi_2\right) \lambda^2 A^2. \qquad
\end{eqnarray}
This concludes the expansion up to second order in $\tau_p H$, i.e.,
in the shortness of the pulse.

\section{Shaping the Pulses}
\subsection{General Discussion}

\subsubsection{Linear Order}

The general idea is that one can shape the pulses such that the
leading deviations in $U_F$ from unity vanish. First, we focus on
the linear order which requires that $\eta^{(1)}$ vanishes.
Hence we have two conditions given by $\eta_{11}=0$ and 
 $\eta_{12}=0$. The relation $v(t)=\partial_t \psi(t)/2$ can be exploited
to integrate the corresponding integrals (\ref{eta11}) and (\ref{eta12}) 
by parts yielding
\begin{subequations}
\begin{eqnarray}
\label{symm1a}
0&=& (\tau_p-\tau_s)\sin\psi(\tau_p)+\tau_s\sin\psi(0)
-\!\!\int_0^{\tau_p}\!\!\!\sin\psi dt\\
0&=& (\tau_p-\tau_s)\cos\psi(\tau_p)+\tau_s\cos\psi(0)
-\!\!\int_0^{\tau_p}\!\!\!\cos\psi dt.\qquad
\label{symm1b}
\end{eqnarray}
\label{linorder}
\end{subequations}
Any pulse which fulfills these two conditions will show only quadratic
deviations from the idealized instantaneous pulse. Note that $\tau_s$
is not a priori fixed and can be considered a free parameter which
can be tuned to fulfill the above conditions. Hence one additional
free parameter is sufficient to obtain a pulse which is
ideal up to linear order. Explicit solutions will be discussed
below.

A last general property worth mentioning is the symmetry of Eqs.\
(\ref{linorder}) under the transformation $v(t)\to v(\tau_p-t)$
which implies $\tau_s\to \tau_p-\tau_s$ and
$\psi(t)\to -\psi(\tau_p-t)$. Thus, if $v(t)$ fulfills Eqs.\
(\ref{linorder}) for $\tau_s$, then $v(\tau_p-t)$ fulfills
them for $\tau_p-\tau_s$ also.

\subsubsection{Quadratic Order}
The requirement $\eta^{(2)}$ in quadratic order adds another three
integrals for $\eta_{21}, \eta_{22}$, and $\eta_{23}$. It is obvious that
integration by parts can also be used to make $v(t)$ disappear in
the expressions for $\eta_{21}, \eta_{22}$, and $\eta_{23}$.
The resulting expressions are
\begin{subequations}
\begin{eqnarray}
\eta_{21} &=& \frac{(\tau_p-\tau_s)^2}{2}\sin\psi(\tau_p)
-\frac{\tau_s^2}{2}\sin\psi(0)
\nonumber\\
&& -\int_0^{\tau_p}\Delta t\sin\psi(t)dt
\label{eta21-partint}
\\
\eta_{22} &=& -\frac{(\tau_p-\tau_s)^2}{2}\cos\psi(\tau_p)
+\frac{\tau_s^2}{2}\cos\psi(0)
\nonumber\\
&& +\int_0^{\tau_p}\Delta t\cos\psi(t)dt
\label{eta22-partint}
\\
\eta_{23} &=& (\tau_p-\tau_s)\tau_s\sin\theta
-\tau_s\int_0^{\tau_p}\sin(\psi(t)-\psi(0)) dt
\nonumber\\
&& -(\tau_p-\tau_s)\int_0^{\tau_p}\sin(\psi(\tau_p)-\psi(t)) dt
\nonumber\\
&&
+\frac{1}{2} \iint_0^{\tau_p}
\sin(\psi_1-\psi_2)\text{sgn}(t_1-t_2)dt_1dt_2,\qquad
\end{eqnarray}
\end{subequations}
where we used $\theta = \psi(\tau_p)-\psi(0)$.

In view of the five expressions to be set to zero, one needs at least five 
free parameters including $\tau_s$, in order to obtain pulses
which show only cubic deviations from the idealized instantaneous pulse.
One way to approach the solution of the above discussed
conditions is to consider symmetric pulses aiming at $\tau_s=\tau_p/2$.
Then $\psi(t)$ is an odd function about $\tau_s$ and the coefficients
$\eta_{11}$ and $\eta_{22}$ vanish by antisymmetry. 
Then only three conditions remain to be solved. But the shape of
the pulse is already fixed to some extent by its symmetry.

A closer inspection reveals that $\pi$ pulses \emph{cannot}
be corrected in second order. This statement is proven
rigorously by the following consideration. We combine the two
real Eqs.\ (\ref{eta21-partint}) and (\ref{eta22-partint}) to the complex
equation
\begin{equation}
0 =\tau_s^2e^{i\psi(0)} - (\tau_p-\tau_s)^2e^{i\psi(\tau_p)}
+2\int_0^{\tau_p}\Delta t e^{i\psi(t)} dt,
\end{equation}
which leads to 
\begin{eqnarray}
\label{exp-factor}\nonumber
\tau_s^2 + (\tau_p-\tau_s)^2 &=& -2\int_0^{\tau_p}\Delta t 
e^{i(\psi(t)-\psi(0))} dt
\\ \nonumber
&\leq & 2 \int_0^{\tau_p} |\Delta t | dt
\\
&\leq &  \tau_s^2 + (\tau_p-\tau_s)^2.
\end{eqnarray}
The equality holds if and only if the exponential factor in the first line of (\ref{exp-factor}) 
changes its value from $-1$ to $1$ abruptly at $\Delta t=0$. This
implies an instantaneous, ideal pulse with infinite amplitude which is
not experimentally realizable. Hence a $\pi$ pulse \emph{cannot}
be corrected in second order. This no-go result is very remarkable
in its general validity.

For $\pi/2$ pulses we found solutions which make $\eta_{21}$ and
$\eta_{22}$ vanish. In spite of our intensive search 
we have not succeeded in finding solutions which additionally make $\eta_{23}$ 
vanish. Hence we \emph{conjecture} that no such solution exists.
But to our knowledge, there is no mathematical proof stating the impossibility
to correct $\pi/2$ pulses in second order.

In this context,
the reader may wonder whether one cannot combine two $\pi/2$ pulses
\begin{figure}
\begin{center}
     \includegraphics[width=\columnwidth]{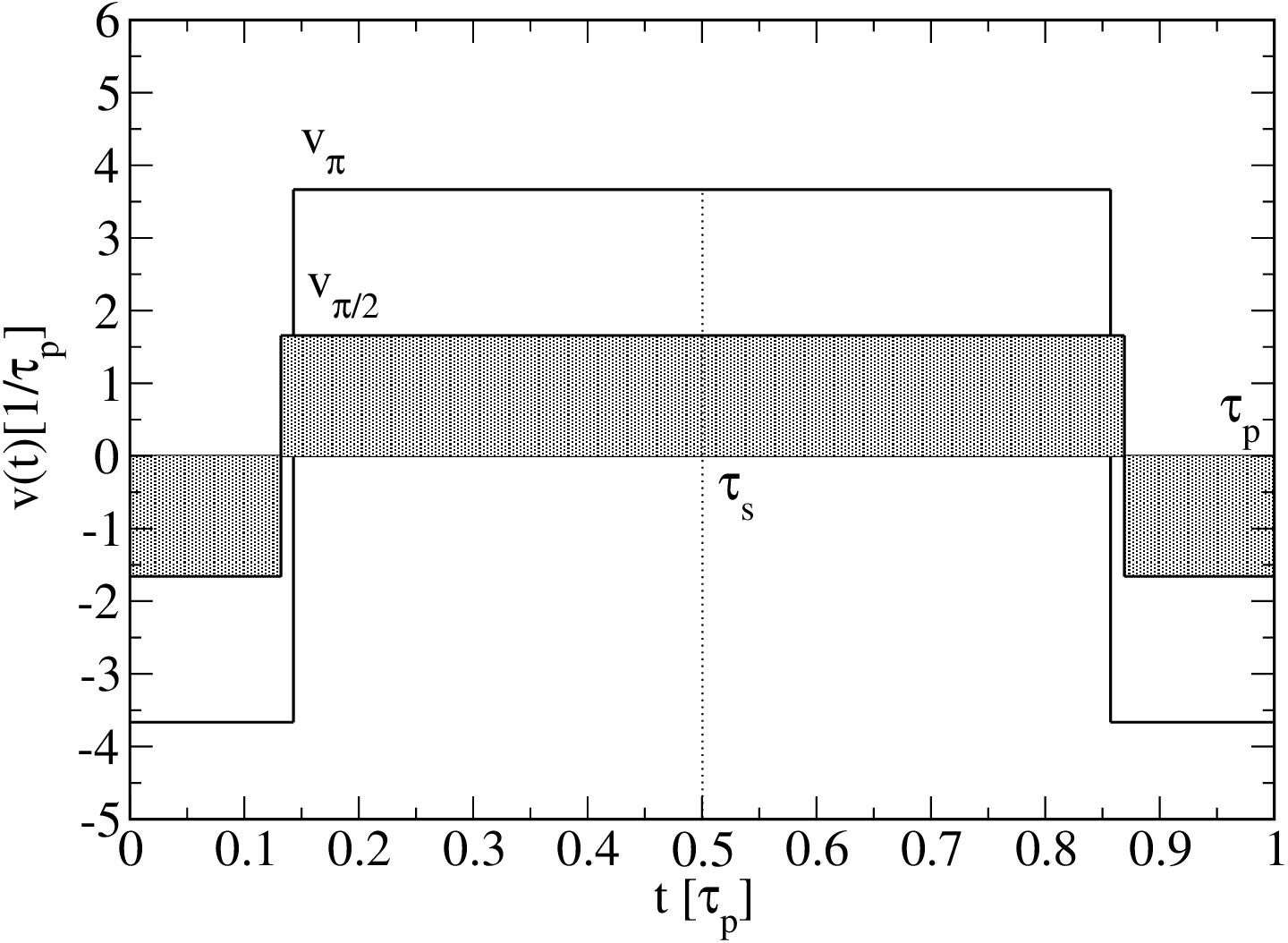}
\end{center}
\caption{Composite $\pi$ and $\pi/2$ pulses with piecewise constant amplitude
 in the symmetric case. In the $\pi$ pulse, the sign changes occur at 
$\tau_1=\tau_p/7$  ($6\tau_p/7$, respectively) and the amplitude is 
$a_\text{max}=7\pi/6 [1/\tau_p]$. In the $\pi/2$ pulse, the sign changes occur 
at $\tau_1=0.13155\tau_p$  ($\tau_p-\tau_1$, respectively) and the amplitude is
$a_\text{max}= 1.65765 [1/\tau_p]$.}
\label{fig-comp-symm}
\end{figure}
to obtain a $\pi$ pulse. But this is indeed not possible within
the present framework. According to \eqref{tot_ev_op} there is
always some decoherent time evolution before and after the 
ideal pulse. Hence no direct combination of pulses with small angles to
pulses with larger angles is possible. But pulse sequences with 
finite time intervals between the pulses as in dynamic decoupling
are well possible 
\cite{viola98,ban98,facch05,khodj05,cappe06,witze07a,yao07,uhrig07}.

\subsubsection{Possibility of $\tau_s=\tau_p$ ?}

We return to the corrections in linear order (\ref{linorder}).
Inspired from discussions with experimentalists in NMR, we pose the question
whether it is possible to shape a pulse such that it corresponds to
an ideal, instantaneous pulse at the \emph{end} of the real pulse, i.e.\
$\tau_s=\tau_p$. If such a pulse existed it could be used to
measure the response of systems at delay times much shorter than
the duration $\tau_p$ of the initial pulse. Clearly, this would
represent a promising experimental tool.

Unfortunately, it is impossible to construct such a pulse. Again,
we can prove this rigorously by inserting $\tau_s=\tau_p$ into
the equations (\ref{linorder}) which we combine to one single
complex equation using $\exp(i\psi)=\cos\psi+i\sin\psi$
\begin{equation}
0=\tau_p \exp(i\psi(0))-\int_0^{\tau_p}\exp(i\psi(t)) dt.
\end{equation}
We find again a rigorous estimate
\begin{eqnarray}
\tau_p &=& \int_0^{\tau_p}\exp\{i[\psi(t)-\psi(0)]\} dt
\nonumber\\
&\leq& \int_0^{\tau_p} 1 dt\ = \ \tau_p
\end{eqnarray}
which is fulfilled if and only if $\psi(t)$ is piecewise
constant with jumps of $2\pi$. But such a $\psi(t)$ represents
\begin{figure}
\begin{center}
     \includegraphics[width=\columnwidth]{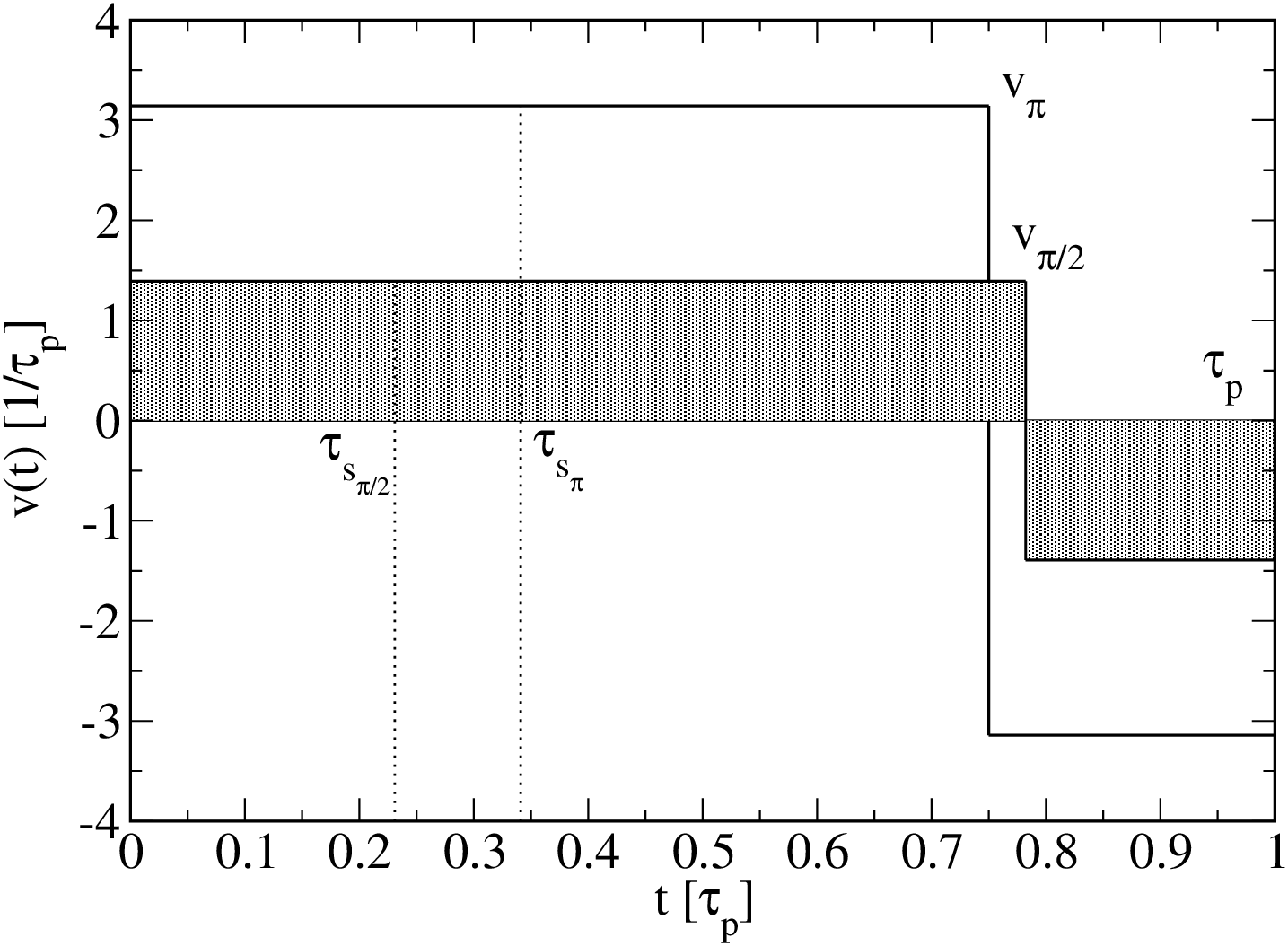}
\end{center}
\caption{Composite $\pi$ and $\pi/2$ pulses with
piecewise constant amplitude made from two elementary
pulses in the asymmetric case. 
For the $\pi$ pulse, the sign change occurs at $\tau_1=3\tau_p/4$, 
the amplitude is $a_\text{max}= \pi [1/\tau_p]$ and
the instant of the equivalent ideal pulse is
$\tau_s=(\tau_p/2)(1-1/\pi)$.
For the $\pi/2$ pulse, $\tau_1=0.78220\tau_p$, the amplitude is 
$a_\text{max}= 1.39156 [1/\tau_p]$ and the instant of the equivalent ideal 
pulse is $\tau_s = 0.23128 \tau_p$.
\label{fig-comp-asym1}}
\end{figure}
via $\partial_t\psi(t)=2v(t)$ instantaneous pulses with infinite
amplitudes. Thus there cannot be a real pulse with bounded amplitude
which is close to an ideal pulse at the end of its duration in linear
order of the bath parameters.

One can only approximate the desired situation by aiming at
a $\tau_s$ close to $\tau_p$. To this end, asymmetric pulses will
be considered below.

\subsection{Solutions}

In view of the above general discussion we focus here only on
the correction in linear order. First, we consider so-called
composite pulses which consist of piecewise constant pulses of
maximally positive or negative amplitude $\pm a_\text{max}$. 
Second, continuously
shaped pulses made from sines and cosines are investigated.
In each class, we look at symmetric pulses with $\tau_s=\tau_p/2$ first 
and then at asymmetric pulses.

\subsubsection{Composite Pulses}

\paragraph{Symmetric Pulses}

For symmetric pulses with $\tau_s=\tau_p/2$ there is only one equation
\begin{figure}
\begin{center}
     \includegraphics[width=\columnwidth]{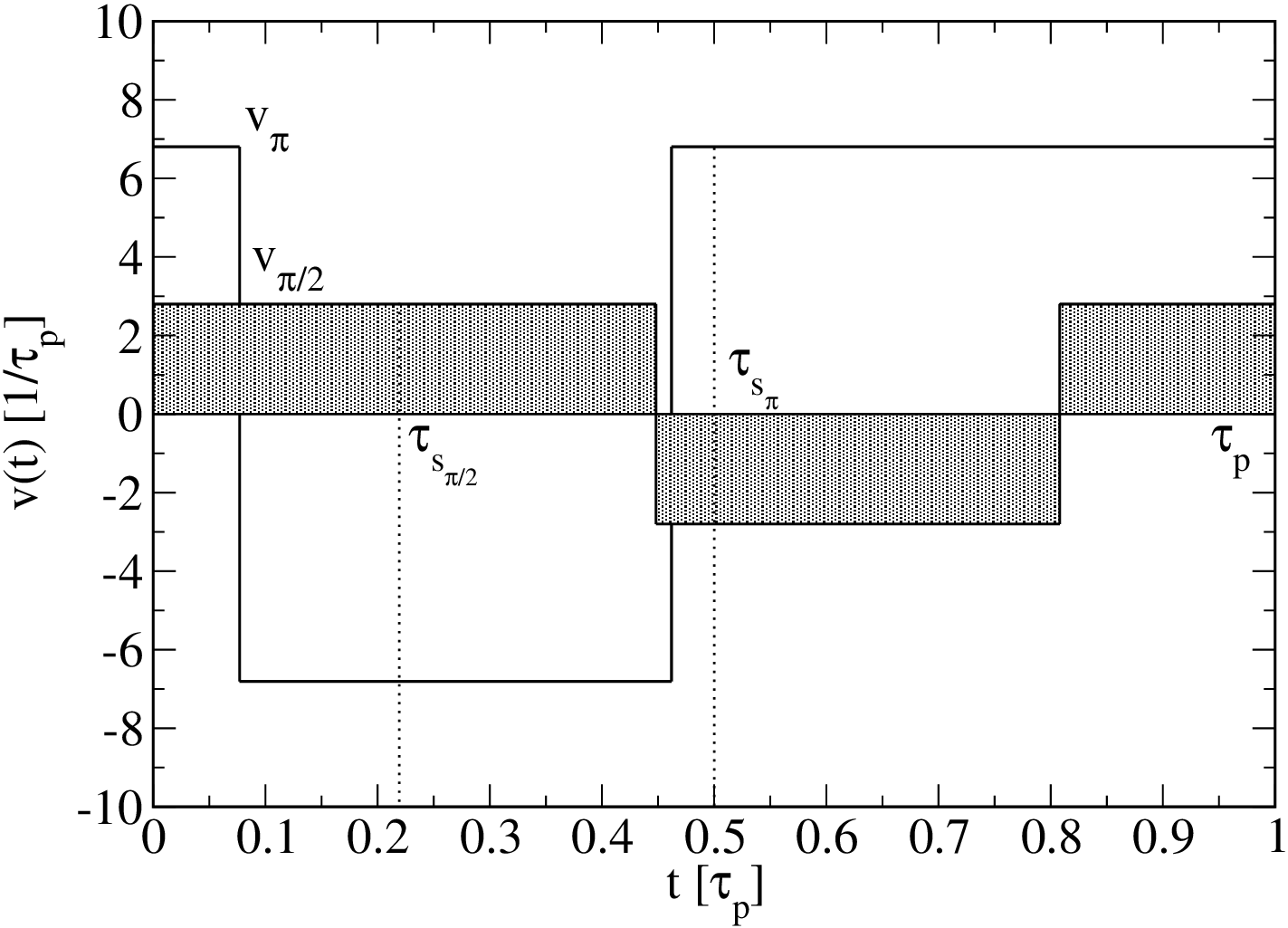}
\end{center}
\caption{Composite $\pi$ and $\pi/2$ pulses with
piecewise constant amplitude made from three elementary
pulses in the asymmetric case. For the $\pi$ pulse, the sign changes occur 
at $\tau_1=\tau_p/13$ and at  $\tau_2=6\tau_p/13$, the amplitude is 
$a_\text{max}= 13\pi/6 [1/\tau_p]$ and the instant of the equivalent ideal 
pulse is $\tau_s=\tau_p/2$. For the $\pi/2$ pulse, the sign changes occur at 
$\tau_1=0.44834\tau_p$ and at $\tau_2=0.80813\tau_p$, the amplitude is 
$a_\text{max}= 2.80074 [1/\tau_p]$ and the instant of the equivalent ideal
pulse is $\tau_s=0.219 \tau_p$.}
\label{fig-comp-asym2}
\end{figure}
to be solved, namely (\ref{symm1b}), because (\ref{symm1a}) is
fulfilled by antisymmetry. We need only one free parameter
and take the instant $\tau_1$, at which the pulse changes first,
to be this free parameter. Another sign change occurs by symmetry
at $t=\tau_p-\tau_1$. Fig.\ \ref{fig-comp-symm} depicts the
solutions for a $\pi$ pulse and for a $\pi/2$ pulse. The parameters
are given in the caption.

It is noteworthy that the symmetric $\pi$ pulse found is exactly
the one that Cummins et al.\ named SCORPSE \cite{cummi00,cummi03}.
Do our results reproduce theirs generally? The answer is ``no'' because
the goals are different. While Cummins et al.\ aim at finding
a pulse shape such that
\begin{equation}
\label{aim-cummins}
U_p(\tau_p,0) \approx U_\theta
\end{equation}
our aim is to approximate as closely as possible
\begin{equation}
\label{aim-pasini}
U_p(\tau_p,0) \approx \exp(-i(\tau_p-\tau_s)H) U_\theta \exp(-i\tau_s H),
\end{equation}
where $U_\theta$ stands for the ideal, $\theta$ pulse evolution operator.
In other words, the authors of the Refs.\ \onlinecite{cummi00,cummi03}
aim at making the imperfections vanish completely while
our aim is more modest: we aim at separating the bath and the
pulse evolution. We are convinced that our aim is more realistic, in 
particular for general dynamic baths 
without static components so that no averaging to zero is possible.
The compensation of the coupling between qubit and bath for longer
storage times shall be done by dynamic decoupling 
\cite{viola98,ban98,facch05,khodj05,cappe06,witze07a,yao07,uhrig07}
on the basis of the optimized pulses discussed here.

But for static couplings $H\propto \sigma_z$ and a $\pi$ pulse
with $\tau_s=\tau_p/2$
both aims (\ref{aim-cummins},\ref{aim-pasini}) happen to coincide
\begin{eqnarray}
\nonumber
e^{-i\tau_p H/2} U_\pi  e^{-i\tau_p H/2}
&=&  U_\pi e^{i\tau_p H/2} e^{-i\tau_p H/2} \\ 
&=&  U_\pi .
\end{eqnarray}
Note that the argument does not require that the pulse is symmetric
\begin{figure}
\begin{center}
     \includegraphics[width=\columnwidth]{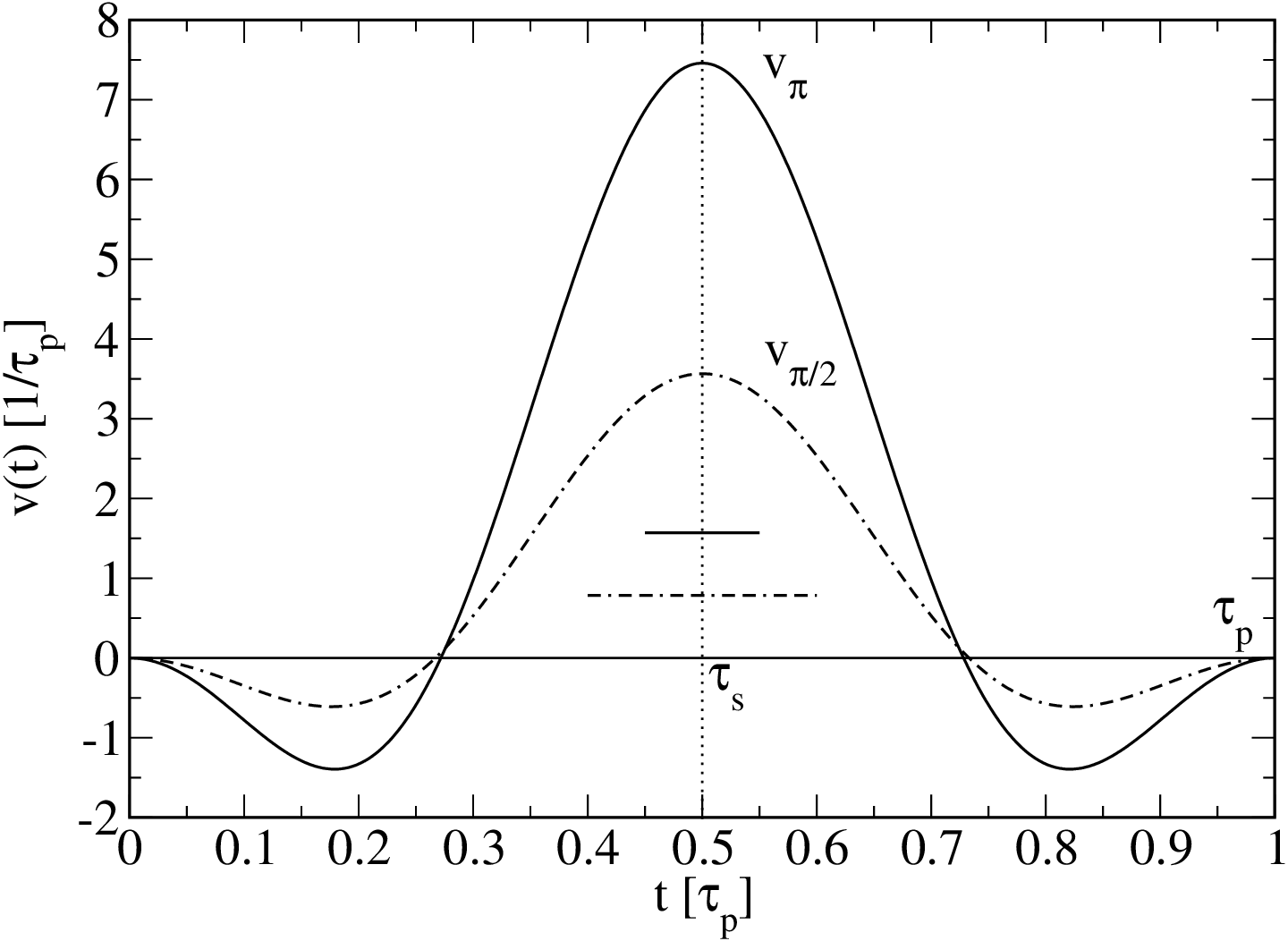}
\end{center}
\caption{Continuous $\pi$ and $\pi/2$ pulses in the symmetric case.  
The paramter $a$ as defined in the ansatz (\ref{ansatz-cont-symm})
takes the value $-3.73002/\tau_p$ for the $\pi$ pulse and the value 
$-1.78259/\tau_p$ for the $\pi/2$ pulse. The horizontal lines 
indicate the amplitude for a totally constant, uncorrected $\pi$ pulse 
(solid line) and $\pi/2$ pulse (dashed-dotted line), respectively.}
\label{fig-cont-symm}
\end{figure}
but only that the splitting time $\tau_s$ is in the middle of the pulse
so that the first half of the evolution under $H$ compensates the
second half. Indeed, the CORPSE pulse \cite{cummi00,cummi03}
fulfills our conditions (\ref{linorder}) as well, see Fig.\ 
\ref{fig-comp-asym2}.

\paragraph{Asymmetric Pulses}

On considering asymmetric pulses there is no reason to restrict oneself
to $\tau_s=\tau_p/2$. Hence there is another free parameter to tune. 
In this way, it is possible to find composite pulses which consist
only of two constant regions.
Fig.\ \ref{fig-comp-asym1} depicts the simplest solutions, but there
can be many more.
For the $\pi$ pulse, for instance, we have found an infinite set
of solutions with the time $\tau_1=(2n+1)\tau_p/(4n)$ at which the jump occurs 
and the amplitude $a_\text{max}=\pi n/\tau_p$ where
$n$ is a positive integer. The corresponding splitting time
reads $\tau_s/\tau_p=1/2+(-1)^n/(2n\pi)$. The solution depicted 
corresponds to $n=1$.

The pulses depicted in Fig.\ \ref{fig-comp-asym1} are the
most asymmetric ones for given maximum amplitude. This is reflected
by the values for $\tau_s$ which are not close to $\tau_p/2$.
If one is looking for a pulse shape which allows the
measurement of a signal as soon as possible after an
ideal pulse the inverted pulses $v(\tau_p-t)$ with
$\tau_s\approx 0.77$ for the $\pi/2$ pulse and 
with $\tau_s\approx 0.66$ for the $\pi$ pulse should be used.

Beyond the composite pulses consisting of
two constant regions we consider also the more commonly
discussed composite pulses consisting of three constant regions.
Fig.\ \ref{fig-comp-asym2} shows generic results.
The $\pi$ pulse depicted is the CORPSE pulse proposed
previously \cite{cummi00,cummi03} which happens to fulfill
also our conditions as explained above. The solution is
arbitrary in the sense that other values for $\tau_s$ are
also possible.

The same is true for the $\pi/2$ pulse for which we have chosen
$\tau_s$ at will. Many other values are also possible. For
$\tau_s=\tau_p/2$, however, we have not found any asymmetric solution.

\subsubsection{Continuous Pulses}

The above composite pulses are advantageous because
they are fairly simple to generate and because they use 
the maximally possible amplitudes most efficiently.
But they are not optimum in view of their frequency selectivity
due to the jumps. It is well-known that the faithful representation
of jumps requires particularly broad frequency bands.
Hence it is important to demonstrate that our conditions (\ref{linorder})
can be also fulfilled by continuous pulses. We choose
ans\"atze inspired from Fourier series which ensure even continuity
of the derivative $v'(t)$.

\paragraph{Symmetric Pulses}

For symmetric pulses our ansatz reads
\begin{equation}
\label{ansatz-cont-symm}
v(t) = \theta/2+a\cos(2\pi t/\tau_p)-(a+\theta/2)\cos(4\pi t/\tau_p).
\end{equation}
The coefficients are chosen such that $v(0)=v(\tau_p)=0$
and $v'(0)=v'(\tau_p)=0$. 
Of course, higher cosine terms could be added. But the constraints
in linear order (\ref{linorder}) can be fulfilled already by
the above ansatz. Fig.\ \ref{fig-cont-symm} displays the resulting
pulse shapes. Note that the instant of the equivalent ideal pulse
is $\tau_s=\tau_p/2$. 
Comparing Fig.\ \ref{fig-comp-symm} and Fig.\ \ref{fig-cont-symm}
the above statement is illustrated that the composite pulses
require only smaller amplitudes. But the continuous pulses
in Fig.\ \ref{fig-cont-symm} are much smoother and, hence,
do not spread so much in frequency space. It depends on the
actual constraints in experiment which kind of pulse is the most
advantageous.

\paragraph{Asymmetric Pulses}

\begin{figure}
\begin{center}
     \includegraphics[width=\columnwidth]{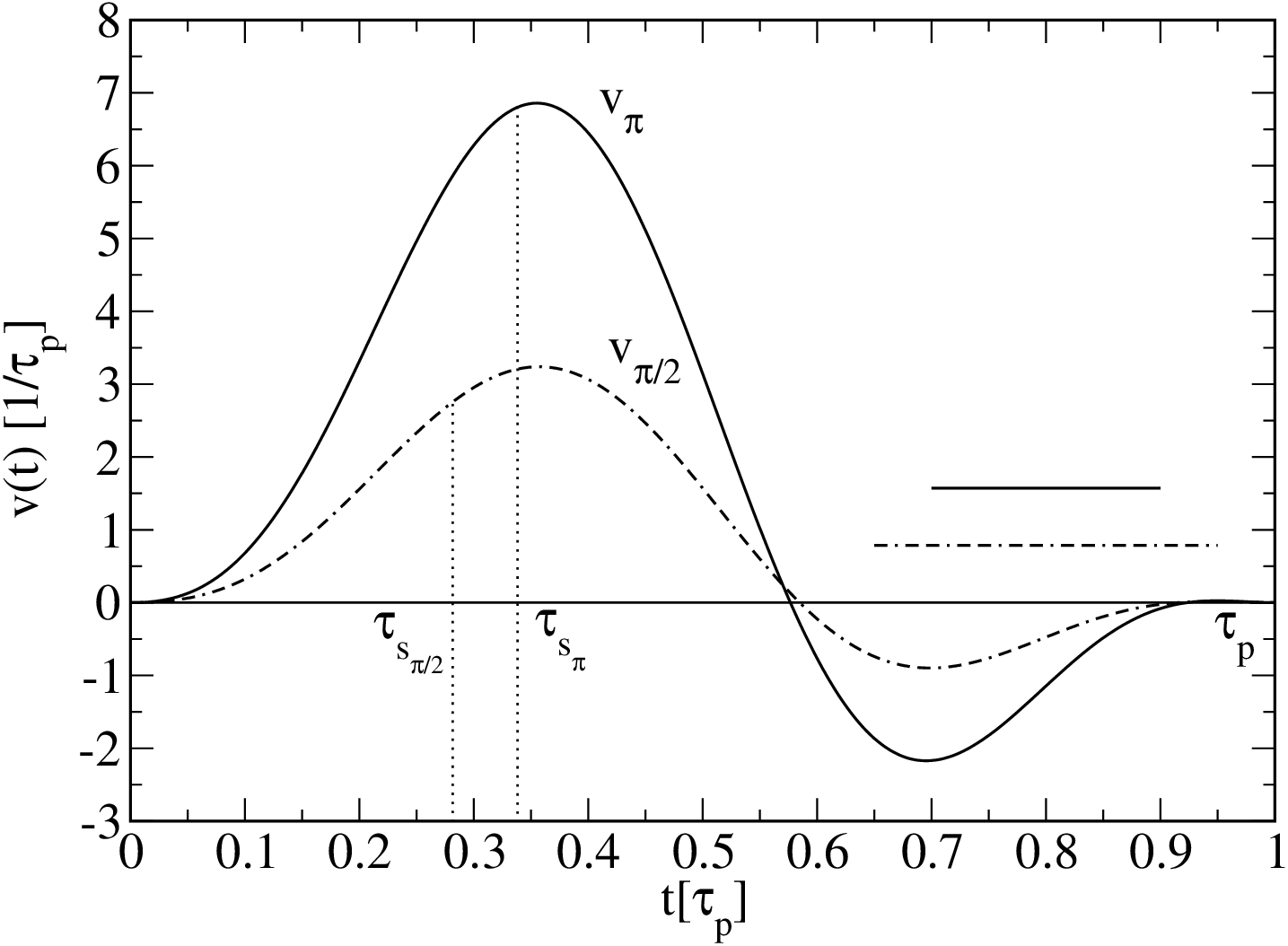}
\end{center}
\caption{Continuous $\pi$ and $\pi/2$ pulses in the asymmetric case. 
For the $\pi$ pulse, the parameter $a$ as defined in the ansatz 
(\ref{ansatz-cont-asym}) takes the value $3.39308/\tau_p$ and the position 
$\tau_s$ of the equivalent ideal pulse takes the value $0.33806\tau_p$.
For the $\pi/2$ pulse, the parameter $a$ is $1.54714/\tau_p$ and 
the position $\tau_s$ of the equivalent ideal pulse is $0.28093\tau_p$. The 
horizontal lines indicate the amplitude for a totally constant, uncorrected
$\pi$ pulse (solid line) and $\pi/2$ pulse (dashed-dotted line), respectively.}
\label{fig-cont-asym}
\end{figure}

For completeness, we also consider an ansatz allowing for asymmetric
continuous pulses
\begin{eqnarray}
\nonumber
v(t) &=& (\theta/2)(1-\cos(2\pi t/\tau_p))\\
&&\qquad +a\sin(2\pi t/\tau_p) -(a/2)\sin(4\pi t/\tau_p).\qquad
\label{ansatz-cont-asym}
\end{eqnarray}
The resulting solutions of the conditions (\ref{linorder}) are plotted
in Fig.\ \ref{fig-cont-asym}. The values of the parameters are
given in the caption. Naturally, the instant of the equivalent
instantaneous pulse is no longer in the middle of the pulse
$\tau_s\neq \tau_p$.

\section{Conclusions}

In this paper, we have investigated under which conditions it is possible
to shape the pulses implementing single qubit gates such that they
correspond to ideal, instantaneous pulses (cf.\ Fig.\ \ref{fig1}).
To this end, we computed the corrections in powers of $H$, the Hamiltonian
containing the coupling to the bath and the dynamics of the bath, to the
equivalence
\begin{equation}
U_p(\tau_p,0) = e^{-i(\tau_p-\tau_s)H} U_\theta e^{-i\tau_s H}
+{\cal O}(H^2).
\end{equation}
The expansion in powers of $H$ is an expansion in the shortness
of the total pulse. No assumption about the nature of the bath
is made so that our results are very generally applicable.

We have derived the explicit expression for the corrections
in linear order in $H$ as well as those quadratic in $H$.
In the evaluation of the expressions found we focussed on the
$\pi$ pulse and the $\pi/2$ pulse. They are by far the most important
ones for all kinds of applications. The $\pi$ pulse is used to
flip spins or qubits. An important application are spin echo
experiments and, more generally, dynamic decoupling of 
two-level systems from their environment
\cite{viola98,ban98,facch05,khodj05,cappe06,witze07a,yao07,uhrig07}.
The $\pi/2$ pulse generates in NMR experiments the precessing
magnetic field and is used in many intricate pulse sequences
to suppress the effect of unwanted interactions \cite{haebe76}.
In the context of quantum information, the  $\pi/2$ pulse
realizes the important Hadamard gate.

We found and showed a multitude of pulse shapes which are 
correct in linear order, i.e.\
\begin{equation}
\label{goal}
U_p(\tau_p,0) = e^{-i(\tau_p-\tau_s)H} U_\theta e^{-i\tau_s H}
+{\cal O}(H^2).
\end{equation}
The pulses can be chosen piecewise constant which requires the 
lowest amplitudes. They imply, however, jumps which deteriorate the
frequency selectivity \cite{vande04}. 
Continuous and continuously differentiable pulses can also be realized and they
have a much better frequency selectivity. Their drawback is that the
maximum amplitude required is larger than for the piecewise
constant pulses.

We noted that our objective (\ref{goal}) happens to coincide 
with the one of earlier work \cite{tycko83,cummi00,cummi03}
for a $\pi$ pulse with $\tau_s=1/2$ applied to a completely static bath,
i.e., a bath without internal dynamics.
One reason for this coincidence is that in leading order the internal
bath dynamics does not play a role. The commutators between
different bath operators occur only in second order.
Hence the corresponding CORPSE and SCORPSE pulse fulfill
also the conditions derived in the present paper.
In general, however, the coincidence does not hold,
i.e.\ if $\tau_s\neq \tau_p/2$ or if the bath is not static 
in the above sense
 or if pulses with angles $\theta$ different from $\pi$ are considered.
Moreover, we emphasize that for dynamic baths 
without static  components,
i.e., baths with internal dynamics,
the aim to make
the coupling between qubit and bath vanish altogether, i.e.\ 
$U_p(\tau_p,0) \approx U_\theta$, cannot be fulfilled for
general angle $\theta\neq\pi$.

Investigating the quadratic order, we could prove that $\pi$ pulses
cannot be corrected in this order. For $\pi/2$ pulses we have
not found any solution in spite of intensive search. Hence we
are led to conjecture that no such solution exists. It would
be interesting to find a mathematical foundation for this conjecture.
Even in linear order, it is not possible to shape the pulse
in such a way that it corresponds to an ideal, instantaneous pulse at the 
very end of the real pulse. This would have been a nice property for
NMR measurements since it would have permitted to measure
directly after a given pulse without further delay. 
One may, however, use asymmetric pulses which correspond
to ideal pulses at about 77\% of the total pulse length.

The message from our findings is that pulses can be shaped 
such that they approximate ideal, instantaneous
pulses. The required pulse shapes have to fulfill rather 
simple analytic integrals so that everyone can fine-tune
his pulse shape easily himself. The optimized $\pi$ pulses
are an excellent starting point for optimized
dynamic decoupling schemes \cite{uhrig07}.

M\"ott\"onen et al.\ \cite{motto06} have posed themselves a similar
question for $\pi$ pulses as we have done. The main differences are that they
investigated classical noise numerically while we tackled 
a fully quantum mechanical model by analytical means. Another 
difference is that they aimed at $U_p(\tau_p,0) \approx U_\pi$
whereas our goal was \emph{not} to make the bath vanish, but to
disentangle it from the actual pulse,
i.e., the coupling between qubit and bath is still present
before and after the idealized pulse.
Of course, in a model of classical noise there are no
bath operators which do not commute so the mere disentanglement
is trivially given in the model studied by M\"ott\"onen et al..
We found that $U_p(\tau_p,0) \approx U_\pi$
and $U_p(\tau_p,0) = e^{-i(\tau_p/2)H} U_\pi e^{-i(\tau_p/2) H}$ coincide
for baths without internal dynamics.
 For these special cases, our pulse shapes
agree with those previously found \cite{cummi00,cummi03,motto06}. 
For instance, the SCORPSE pulse appears to be a very good choice.

Further work should simulate the proposed
pulse shapes in specific models in order to understand
better how important the neglected corrections in second
order really are. It is to be expected that for small
values of $\lambda$ (generic coupling to the bath) and $\omega_b$
(generic frequency of the bath) the corrected pulses are
superior to the simple, uncorrected ones. But if the characteristic
times of the bath are too short, presumably the simple pulses will be
better.

\begin{acknowledgments}
We would like to thank  M.~Lovri\'c, J.~Stolze, and D.~Suter for 
helpful discussions. The financial support in the GK 726 by the DFG
is gratefully acknowledged.
\end{acknowledgments}


\end{document}